\documentclass[twocolumn,floatfix,superscriptaddress,amsmath,showpacs,showkeys,aps,prb]{revtex4}

\usepackage[final]{graphicx}
\usepackage{t1enc}
\usepackage{bm}

\begin{document}

\bibliographystyle{apsrev}

\title{\bf Phase diagram of an Ising model for ultrathin magnetic films:
comparing mean field and Monte Carlo predictions}

\author{Santiago A. Pigh\'{\i}n}
\email{spighin@famaf.unc.edu.ar} \affiliation{Facultad de
Matem\'atica, Astronom\'{\i}a y F\'{\i}sica, Universidad Nacional
de C\'ordoba,
\\ Ciudad Universitaria, 5000 C\'ordoba, Argentina}
\altaffiliation{Members of CONICET, Argentina}

\author{Sergio A. Cannas}
\email{cannas@famaf.unc.edu.ar} \affiliation{Facultad de
Matem\'atica, Astronom\'{\i}a y F\'{\i}sica, Universidad Nacional
de C\'ordoba, \\ Ciudad Universitaria, 5000 C\'ordoba, Argentina}
\altaffiliation{Members of CONICET, Argentina}

\date{\today}

\begin{abstract}
We study the critical properties of a two--dimensional Ising model
with competing ferromagnetic exchange and dipolar interactions,
which models an ultra-thin magnetic film with high out--of--plane
anisotropy in the monolayer limit. In this work we present a
detailed calculation of the $(\delta,T)$ phase diagram, $\delta$
being the ratio between exchange and dipolar interactions
intensities.  We compare the results of both mean field
approximation and Monte Carlo numerical simulations in the region
of low values of $\delta$, identifying the presence of a recently
detected phase with nematic order in different parts of the phase
diagram, besides the well known striped and tetragonal liquid
phases. We also found that, in the regions of the phase diagram
where Monte Carlo simulations display nematic order, the mean
field approximation predicts hybrid solutions composed by stripes
of different widths. Another remarkable qualitative difference
between both calculations is the absence, in this region of the
Monte Carlo phase diagram, of the temperature dependency of the
equilibrium stripe width predicted by the mean field
approximation.

\end{abstract}


\maketitle

\section{Introduction}

Many ultrathin magnetic films, like e.g. Co/Cu, Co/Au, Fe/Cu,
undergo a reorientation transition at a temperature $T_R$; for
temperatures below $T_R$ the spins align preferentially in a
direction perpendicular to the film, while above  $T_R$ they align
in a magnetized state parallel to the plane of the
film~\cite{AlStBi1990,VaStMaPiPoPe2000,DeMaWh2000}. This
reorientation transition is due to the competition between the
in-plane part of the dipolar interaction and the surface
anisotropy~\cite{Po1998}. Furthermore, in the range of
temperatures where the magnetization points out of the plane, the
competition between exchange and dipolar interaction causes the
global magnetization to be effectively zero but instead striped
magnetic domain patterns
emerge\cite{VaStMaPiPoPe2000,PoVaPe2003,DeMaWh2000}. In the limit
of a monolayer film, the following dimensionless Ising Hamiltonian
emerges as a minimal model to describe many of the relevant
physical properties of those materials\cite{DeMaWh2000}:

\begin{equation}
{\cal H}= - \delta \sum_{<i,j>} S_i S_j + \sum_{(i,j)} \frac{S_i
S_j}{r^3_{ij}} \label{Hamilton1}
\end{equation}

\noindent where $\delta$ stands for the ratio between the exchange
$J_0>0$ and the dipolar $J_d>0$ interactions parameters, i.e.,
$\delta = J_0/J_d$. The first sum runs over all pairs of nearest
neighbor Ising spins $S_i = \pm 1$ in a square lattice and the
second one over all distinct pairs of spins of the lattice (every
pair is counted once); $r_{ij}$ is the distance, measured in
crystal units, between sites $i$ and $j$ and the energy is
measured in units of $J_d$. In spite of intense theoretical
work\cite{CzVi1989,DeMaWh2000,MaWhRoDe1995,BoMaWhDe1995,StSi2000,GlTaCa2002,CaGlTa2004,CaStTa2004,CaDaRaRe2004,CaMiStTa2006,RaReTa2006},
there are still many important open questions regarding the
critical properties of this model. A detailed understanding of
those critical properties is the cornerstone of the theoretical
framework needed to explain complex phenomena
 in ultrathin magnetic films, such as the recently observed inverse
 transition in Fe on Cu films\cite{PoVaPe2003}.

The main difficulties to analyze the critical properties of this
model are related to the long range character of the dipolar
interactions, which combined with the frustration derived from the
competition between interactions, adds to any theoretical approach
an extra degree of complexity. Then, even the simplest approach,
namely mean field approximation (MF), leads to an infinite number
of coupled equations that, except for some particular situations
cannot be solved exactly. For instance, in an early work Czech and
Villain\cite{CzVi1989} derived an exact expression for the MF
critical temperature between the disordered and the modulated
(striped) phases; however, for sub--critical temperatures the
determination of any property must rely on numerical solutions of
the mean field equations or further ansatz  has to be
introduced\cite{PoVaPe2006} to obtain {\it approximated} solutions
of the MF equations. An example is the temperature dependency of
the equilibrium stripe width; being experimentally
accessible\cite{PoVaPe2006}, reliable theoretical predictions of
this property could be very important to understand the basic
mechanisms behind the complex behavior observed in these
materials. Though MF is a powerful theoretical tool, it is known
that, even when the corresponding equations can be solved exactly,
neglecting fluctuations can introduce qualitative changes in the
critical behavior. Therefore, it is important to compare MF
predictions with those obtained by other methods, in order to
establish the limits of validity of the approximation.

 A natural way to check the mean field predictions is
to contrast them with Monte Carlo (MC) simulations. However, the
long range order nature of the dipolar interactions makes it very
difficult to simulate large system sizes. The ground state stripe
width $h$ is the natural length scale in these problems. Hence, in
order to avoid strong finite size effects, the simulations must be
carried out for system sizes $L \gg h$; this restricts the
simulations to situations in which $h$ is much smaller than the
experimentally observed values (typical values of the stripe width
in Fe on Cu films, for instance, are of the
order\cite{WoWuKiScDoOwWuJiZhQi2005,PoVaPe2006} of $1\, \mu m$,
which corresponds roughly to $h=4000$ lattice constants). Since
the ground state value of $h$ increases exponentially
with\cite{YaGy1988} $\delta$, the values of $\delta$ available for
simulations are about of one order of magnitude smaller then the
realistic values.

 In this work we carry out a detailed analysis
of the equilibrium phase diagram of this model in the $(\delta,T)$
space for low values of $\delta$, i.e., for $0 \leq \delta \leq
4$, which corresponds to stripe widths $h \leq 7$. By extending
the Czech and Villain\cite{CzVi1989} MF approach to the low
temperature region of the phase diagram, we performed in section
\ref{mean field} a detailed analysis of the different transition
lines between striped states, by solving numerically the MF
equations. Those results are compared in section \ref{MC} with MC
simulations that refine previous
results\cite{MaWhRoDe1995,GlTaCa2002,CaGlTa2004,CaDaRaRe2004,CaMiStTa2006,RaReTa2006}
and extend them to other regions of the phase diagram. Our results
show that, at least in the analyzed region of the phase diagram,
several discrepancies are observed between both phase diagrams,
which are discussed in section \ref{disc}.

\section{Mean field phase diagram}
\label{mean field}

The Hamiltonian (\ref{Hamilton1}) can be rewritten as

\begin{equation}
  \label{HamiltonianJ}
    {\cal H} = -\frac{1}{2} \sum_{i,j} J_{ij} \, S_i S_j
\end{equation}

\noindent where

\begin{equation}
J_{ij} = \left\{
\begin{array}{ll}
\delta-1          & \mbox{\,\, if \, $i,j$ are nearest neighbors} \\
0                  & \mbox{\,\, if \, $i=j$}  \\
-\frac{1}{r_{ij}^3} & \mbox{\,\, otherwise}
\end{array} \right.
\end{equation}

A straightforward way to derive a mean field theory for this
Hamiltonian is the usage of the variational MF free energy per
particle\cite{ChLu1995}:

\begin{equation}
f_{mf} = \frac{1}{N} \left< {\cal H} \right>_\rho + \frac{1}{\beta
N} \left< \ln{\rho} \right>_\rho
\end{equation}

\noindent where $N=L \times L$ is the system size, we have taken
$k_B=1$ and the averages are taken using the independent particle
density matrix $\rho = \prod_i \rho_i$; the one particle density
matrices are subjected to the constraints:

\[ \sum_{S_i=\pm 1} \rho_i =1  \;\;\;\;\;\; \sum_{S_i=\pm 1} S_i \rho_i
=m_i. \]

\noindent Using the local order parameters $m_i$ as variational
parameters we obtain the free energy functional

\begin{widetext}
\begin{equation}
f_{mf}[\{ m_i\}]= -\frac{1}{2\, N} \sum_{i,j} J_{ij} \, m_i m_j +
\frac{1}{2 \beta N} \sum_i \left[(1+m_i) \ln{(1+m_i)} + (1-m_i)
\ln{(1-m_i)}\right] \label{fmf}
\end{equation}
\end{widetext}

Minimizing Eq.(\ref{fmf}) respect to the order parameters $m_i$
leads to the set of MF equations

\begin{equation}
m_i =  tanh\left(\beta \sum_{j=1}^N J_{ij} m_j \right) \;\;\;
i=1,\ldots,N \label{mfeq1}
\end{equation}

Assuming periodic boundary conditions, we introduce the Fourier
transforms

\begin{equation}
m_i = \frac{1}{\sqrt{N}} \sum_{\vec{k}} \hat{m}_{\vec{k}} \, e^{i
\vec{k}. \vec{r}_i} \label{fourier}
\end{equation}

\begin{eqnarray}
\hat{J}(\vec{k}) &=& \sum_i J_{0i} \, e^{-i \vec{k}. (\vec{r}_i -
\vec{r}_0)}\nonumber \\
&=& 2\, \delta (\cos{k_x} + \cos{k_y}) - \sum_i \frac{1}{r_{ij}^3}
\cos{(\vec{k}.\vec{r}_{ij})}
\label{fourierJ}
\end{eqnarray}

\noindent where $\vec{r}_i$ is the position vector of site $i$,
$\vec{r}_{ij}\equiv \vec{r}_i-\vec{r}_j$,
$\hat{m}_{-\vec{k}}=\hat{m}_{\vec{k}}^*$ and the wave vectors
$\vec{k}$ are restricted to the first Brillouin zone. Expanding
the logarithms, Eq.(\ref{fmf}) can be rewritten as

\begin{widetext}
\begin{equation}
  f_{mf} =  \frac{1}{2\, N} \sum_{\vec{k}}  \left(T -\hat{J}(\vec{k})  \right)
  \left| \hat{m}_{\vec{k}}\right|^2  + \frac{1}{\beta N} \sum_i
  \sum_{j=2}^{\infty}
  \left( \frac{1}{2j-1} - \frac{1}{2j} \right) m_i^{2j}
  \label{freefunctional}
\end{equation}
\end{widetext}

\noindent which has the form of a Landau expansion. From this
expression it is immediate that a second order  transition between
the disordered phase ,$\hat{m}_{\vec{k}}\equiv 0$ $\forall
\vec{k}$, and an ordered phase, with non--zero order parameters,
happens at the critical temperature\cite{CzVi1989}:

\begin{equation}
  \label{Tc}
  T_c = \max_{\vec{k}} \, \hat{J}(\vec{k})
\end{equation}

\noindent We calculated $T_c(\delta)$ by solving Eq.(\ref{Tc})
numerically.

Eqs.(\ref{mfeq1}) can now be written as

\begin{equation}
  \label{mfeq2}
  m_i =  \tanh \left\{ \frac{\beta}{\sqrt{N}}
  \sum_{\vec{k}} \hat{m}_{\vec{k}} \, e^{i \vec{k}. \vec{r}_i}
  \hat{J}({\vec{k}})
    \right\} \;\; i=1,\ldots,N.
\end{equation}

\begin{figure}
\begin{center}
\includegraphics[scale=0.3,angle=-90]{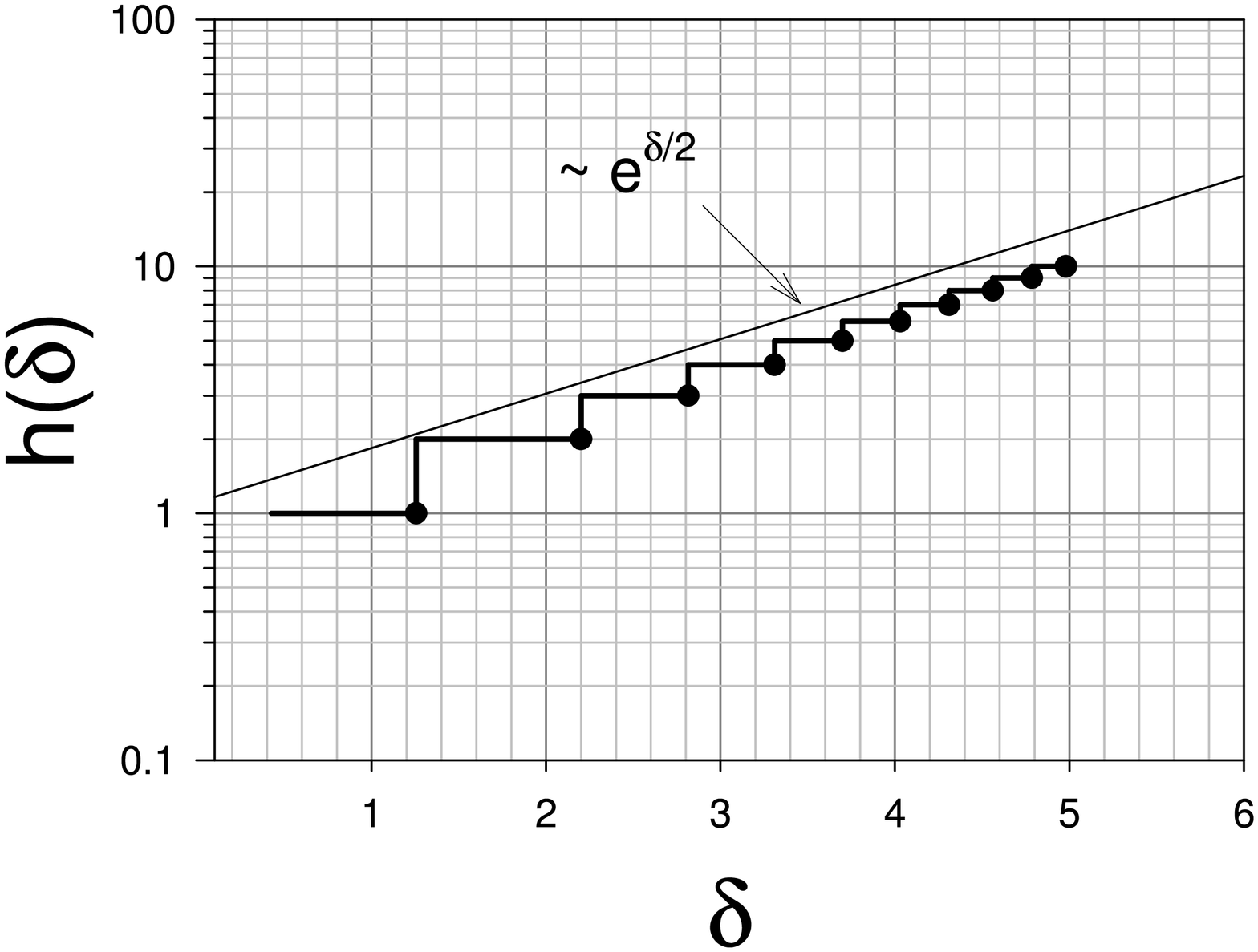}
\caption{\label{fig1} Equilibrium stripe width  as a function of
$\delta$ at $T=0$. Filled circles indicate the values of delta at
which two adjacent striped configurations take the same energy.}
\end{center}
\end{figure}

\noindent We analyzed numerically the solutions of
Eqs.(\ref{mfeq2}) for temperatures $T< T_c$ and $0 < \delta \leq
4$. In particular, we analyzed the solutions that share the
symmetries of the different ground states, namely,
antiferromagnetic and striped solutions. For $\delta < 0.425$ the
ground state is antiferromagnetic\cite{MaWhRoDe1995}, while for
$\delta > 0.425$ the ground state is composed by stripes of width
$h(\delta)$. For large values of $\delta$ we
have\cite{MaWhRoDe1995} $h(\delta) \sim e^{\delta/2}$; for small
values of $\delta$ the equilibrium values of $h$ can be easily
evaluated numerically by comparing the energies of different
striped configurations of increasing finite system sizes (they
converge very quickly). $h(\delta)$ is shown in Fig.\ref{fig1},
where we see that it attains the asymptotic exponential behavior
for rather small values of $\delta$.

At low but finite temperatures, the local magnetization inside the
stripes decrease, i.e., $|m_i|<1$. Let us consider, for instance,
a vertical striped state of width $h$. We demand the solutions of
Eqs.(\ref{mfeq2}) to satisfy the conditions:
$m_{(x+h,y)}=-m_{(x,y)}$ $\forall x,y$ and $m_{(x,y)}=m_{(x,y')}$
$\forall y,y'$. This restricts the harmonics in Eq.(\ref{fourier})
to those satisfying $(k_x,k_y)=(\pm (2l+1)\pi/h,0)$, with $l$ an
integer such that $2l+1 \leq h$. For instance, for $h=1$ we only
have $k_x=\pi$; for $h=2$ we have $k_x=\pm \pi/2$; for $h=3$ we
have $k_x=\pm \pi/3, \pi$; etc. In other words, for a stripe
solution of width $h$ we have $h$ independent complex amplitudes
$\hat{m}_{\vec{k}}$. In order to obtain pure real solutions we
must impose $\hat{m}_{\vec{k}}^*=\hat{m}_{-\vec{k}}$. Replacing
those conditions into Eqs.(\ref{fourier}) and (\ref{mfeq2}) leads
to a set of $h$ non-linear algebraic equations for the amplitudes
that can be solved numerically. To solve those equations we must
evaluate $\hat{J}(k_x,0)$ from Eq.(\ref{fourierJ}). A suitable
approximation for that function is (see Appendix \ref{AppA})

\begin{equation}
  \label{Jkx}
  \hat{J}(k_x,0) \approx
      2 \, \delta \, (\cos k_x + 1) - k_x^2 + 2 \pi |k_x| - \frac{2
      \pi^2}{3} - 2 \, \zeta(3)
\end{equation}

\noindent where $\zeta(x)$ is the Riemann zeta-function. For the
antiferromagnetic solution $m_{(x,y)}=m_0 (-1)^{x+y}$ we have to
compute

\begin{equation}
  \hat{J}(\pi,\pi) =
    -4 \, \delta + 3  \, \zeta(3) - 4 \, \sum_{x=1}^{\infty} (-1)^x \sum_{y=1}^{\infty}
    \frac{(-1)^y}{(x^2+y^2)^{\frac{3}{2}}}
    \label{Jpipi}
\end{equation}

\noindent where the last term is calculated numerically. We
calculated the MF stripe solutions for $h=1,\ldots,6$ for a wide
range of values of $(\delta,T< T_c)$. To discriminate whether they
are actually minima  we analyzed the second derivatives of the
free energy. For every value of $h$ we first analyzed the
stability of the solutions, that is, for every value of $\delta$
we calculated the temperature $T_s(\delta)$ above which non
trivial solutions of the above described type cease to exist. This
can be done  by linearizing the corresponding set of equations
around $\hat{m}_{\vec{k}}=0$ and demanding the condition of
non--trivial solution, i.e., zero determinant of the linearized
equations; this leads to the expression

\begin{equation}
  T_s(\delta,h) = \hat{J}\left(\frac{\pi}{h},0\right)
\end{equation}

\begin{figure}
\begin{center}
\includegraphics[scale=0.3,angle=0]{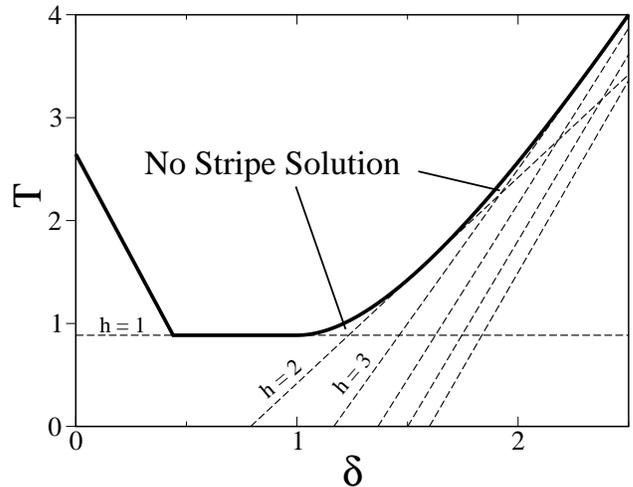}
\caption{\label{fig2} Stability lines $T_s(\delta,h)$ for the MF
striped solutions at finite temperatures (dashed). Full line:
critical temperature $T_c(\delta)$.}
\end{center}
\end{figure}

\begin{figure}
\begin{center}
\includegraphics[scale=0.3,angle=0]{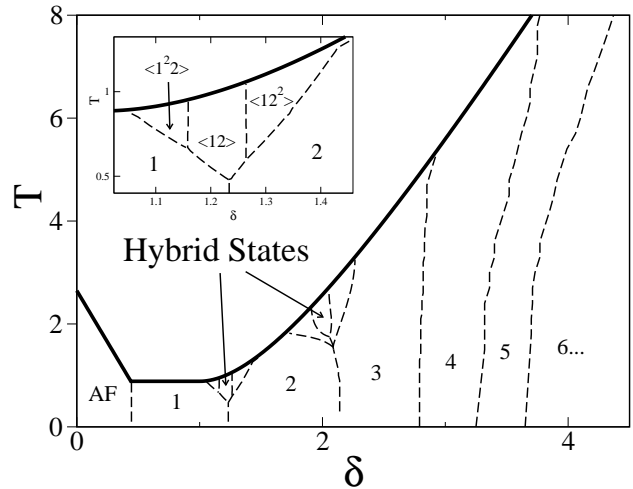}
\caption{\label{fig3} Mean field phase diagram. The numbers in the
low temperature region indicate the equilibrium stripe width $h$
of the phases. Dashed lines correspond to first order phase
transition between ordered phases,; the full line corresponds to
the critical temperature $T_c(\delta)$. Inset: zoom of the hybrid
region near the boundary between the striped phases $h=1$ and
$h=2$.}
\end{center}
\end{figure}

\noindent The stability lines are shown in Fig.\ref{fig2},
together with $T_c(\delta)$. It can be seen that for large values
of $\delta$ the stability lines accumulate near the order-disorder
transition line $T_c(\delta)$, implying an increasingly large
number of metastable states as $\delta$ increases. Another
remarkable fact, is the presence of regions near $T_c(\delta)$
where no striped solutions exists (see Fig.\ref{fig2}). In those
regions another type solutions appear, which are composed by
parallel ferromagnetic stripes of different widths. We called them
hybrid states. The hybrid states, which we denote by $<h_1^{n_1}\,
h_2^{n_2}\, \ldots\,h_l^{n_l}>$, following the notation of Selke
and Fisher \cite{SeFi1979} for the axial next-nearest-neighbor
Ising model (ANNNI), consists in the periodic repetition of a
fundamental pattern composed by $n_1$ stripes of width $h_1$ (with
opposite orientation), followed by $n_2$ stripes of width $h_2$
and so on. The regions where the hybrid states appear are shown in
the MF phase diagram presented in Fig.\ref{fig3}. The boundaries
between ordered phases (corresponding to first order transitions)
where determined by comparing the free energies of the different
solutions (striped and hybrid) using Eq.\ref{fmf}; they are shown
by dashed lines in Fig.\ref{fig3}. We found that the hybrid states
appear through a sort of branching process near the boundary
between two stable striped solutions as the temperature approaches
$T_c$ from below. For instance, the transition line between the
striped phases $h=1$ and $h=2$ ends in a triple point where a
stable phase  $<1\, 2>$ appears between them; as we increase
$\delta$ the transition line between the striped phase $h=2$ and
the hybrid one $<1\, 2>$ bifurcates in a new triple point giving
rise to the appearance of a $<1\, 2^2>$ phase between the $<1\,
2>$ and the $h=2$ and so on (see inset of Fig.\ref{fig3}). As the
temperature increases more complicated hybrid states proliferate
(we just show a few of them in Fig.\ref{fig3} as an example), in a
completely analogous way as in  a related model, namely, the three
dimensional Ising model with competing short range ferromagnetic
interactions and long range Coulomb interactions\cite{GrTaVi2000}.
The MF phase diagram of that model is very similar to that of the
present one, the striped states being replaced by lamellar ones.

Finally, we found evidences that the proliferation of hybrid
states also happens near the boundary between striped phases with
larger widths (for instance, 3 and 4), but they appear very close
to $T_c$ and the computational effort needed to obtain an accurate
estimation of the phase boundaries becomes very high.

\section{Monte Carlo phase diagram}
\label{MC}

Different parts of the phase diagram of this model were analyzed
by different authors using MC numerical simulations, for small
values of $\delta$ and small system
sizes\cite{MaWhRoDe1995,BoMaWhDe1995,GlTaCa2002,CaStTa2004,CaMiStTa2006,RaReTa2006}.
In this section we extend those results to other parts of the
phase diagram and to larger system sizes (in some cases), in order
to obtain a complete description  of the small-$\delta$ phase
diagram that can be compared with the MF phase diagram.

The MC results were obtained using heat bath dynamics on $N = L
\times L$ square lattices with periodic boundary conditions (Ewald
sums were used to handle it\cite{DeMaWh2000}). We analyzed the
equilibrium behavior of different quantities for system sizes
running from $L=24$ to $L=84$; the maximum size used for each
quantity  were chosen according to the associated computational
effort.

The first quantity we calculated was the order--disorder
transition temperature as a function of $\delta$, which we called
$T_c^{(2)}(\delta)$ (analog to $T_c(\delta)$ in the MF case). This
quantity was determined by means of  the specific heat

\begin{equation}
C(T) = \frac{1}{NT^2} \left( \left< H^2 \right> -\left< H
\right>^2\right) \label{heat}
\end{equation}

\noindent where $\left< \cdots \right>$ stands for a thermal
average. For some values of $\delta$ we also calculated the fourth
order cumulant

\begin{equation}
V(T) = 1- \frac{\left< H^4 \right>}{3\left< H^2 \right>^2}
\label{cumulant}
\end{equation}

\noindent to characterize the order of the phase transition.

\begin{figure}
\begin{center}
\includegraphics[scale=0.3,angle=0]{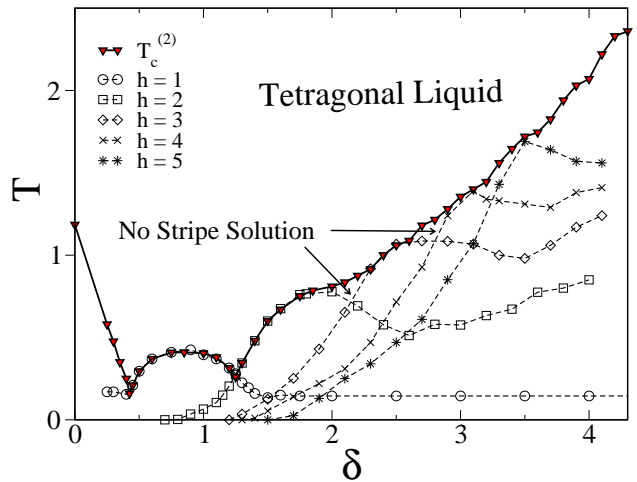}
\caption{\label{fig4} (Color on-line) Order-disorder transition
temperature $T_c^{(2)}(\delta)$ and striped stability lines
$T_s(\delta)$ for different values of $h$ (see text for details).}
\end{center}
\end{figure}

 At intermediate high temperatures (close to the order disorder
transition and above it) this system presents a partially
disordered phase with broken orientational order called {\it
tetragonal
liquid}\cite{BoMaWhDe1995,DeMaWh2000,CaStTa2004,CaDaRaRe2004,CaMiStTa2006}.
It is characterized by domains of stripes with mutually
perpendicular orientations forming a kind of labyrinthine
structure. At higher temperatures these domains collapse and the
system crosses over continuously to a completely disordered phase
(paramagnetic), without a sharp phase transition between
them\cite{DeMaWh2000}. While the existence of the tetragonal
liquid phase has been clearly established by MC simulations, it is
completely absent in the MF approximation (we discuss this point
in section \ref{disc}). MC simulations also showed
recently\cite{CaMiStTa2006} that for $\delta=2$ an intermediate
phase with nematic order is present between the tetragonal liquid
and the striped phase. The nematic phase is characterized by
positional disorder and long range orientational order and is
consistent with one of the two possible scenarios predicted by a
continuum approximation for ultrathin magnetic films in
Refs.\cite{KaPo1993b,AbKaPoSa1995}. The presence of the nematic
phase is reflected (among several manifestations) in the
appearance of two distinct maxima at different temperatures in the
specific heat, associated with the stripe--nematic and the
nematic--tetragonal liquid phase transitions respectively. On the
other hand, it was shown that for $\delta=1$ the specific heat
present a unique maximum, consistent with a direct transition from
the tetragonal liquid to the striped phase\cite{CaMiStTa2006},
suggesting that the nematic phase is only present for some range
of values of $\delta$.  We will call $T_c^{(2)}$ the temperature
of the high temperature peak of the specific heat, whenever it
presents two peaks, or the temperature of the unique peak if only
one is present (for the system sizes considered and between the
precision of the calculation). We will call $T_c^{(1)}$ the
temperature of the low temperature peak of the specific heat, when
it presents two peaks. While the calculation of $T_c^{(2)}$ is
relatively easy, the calculation of $T_c^{(1)}$ is much more
complicated and subtle, as it will be discussed below. $T_c^{(2)}$
was calculated for different values of $0 \leq \delta \leq 4.2$
using the following simulation protocol: for each value of
$\delta$ we let first the system thermalize at a high enough
temperature (such that it is in the disordered phase) during
$10^5$ Monte Carlo Steps (MS); after that, we calculated the
specific heat for decreasing temperatures, down to a temperature
well below the transition region. For every temperature we took
the final configuration of the previous one and discarded the
first $2 \times 10^4$ MCS for thermalization and averaged over
$10^5$ MCS. Every curve was averaged then over $40$ independent
runs. This calculation was performed for system sizes $L \sim 50$
for all the values of $\delta$, in order to make the finite size
bias comparable (for every value of $\delta$ we choose the closer
value of $L$ commensurated with the modulation $2h$ of the
corresponding ground state width). The results are shown by
triangles joined by a continuous line in Fig.\ref{fig4}. The
 order of the associated phase transition will be
discussed in section \ref{disc}.

\begin{figure}
\begin{center}
\includegraphics[scale=0.3,angle=0]{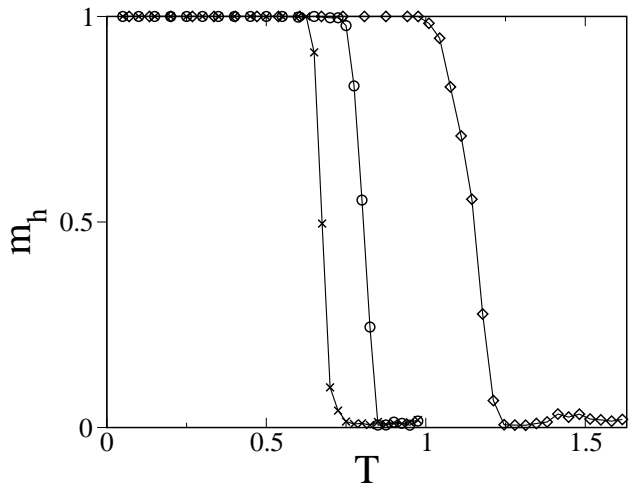}
\caption{\label{fig5} Staggered magnetization for different
$\delta$ and $h$ values:
         $h=2$, $\delta = 3.4$ and $L=60$ ($\times$);
         $h=3$, $\delta = 4.1$,  and $L=48$ ($\diamond$);
         $h=5$, $\delta = 2.9$ and $L=40$ ($\circ$).}
\end{center}
\end{figure}

We next calculated the stability of the striped phases in
different parts of the phase diagram. It was shown in
Ref.\cite{GlTaCa2002} that the striped phases can remain in a
meta-stable state for values of $\delta$ such that the equilibrium
ground state width corresponds to a different stripe width.
Following the same procedure as in Ref.\cite{GlTaCa2002} we
analyzed the striped staggered magnetization

\begin{equation}
  m_h =  \frac{1}{N} \left< \left| \sum_{x,y} \,
  (-1)^{f_h(x)} \, S_{x,y} \, \right| \right>
\end{equation}

\noindent where\cite{DeMaWh2000} $f_h(x) = (i-mod(x,h))/h$ and
$x,y=1,\ldots,N$. This quantity takes the value one in a
completely ordered vertical striped state of width $h$ and zero in
a disordered state (paramagnetic, tetragonal or nematic). Starting
from an initially ordered vertical striped state at zero
temperature we increase the temperature up to high temperatures,
averaging $m_h$ at every step and using  a similar procedure as
for the calculation on $T_c^{(2)}$, but averaging over $10^6$ MCS
for every temperature to diminish metastability effects (see
discussion below). We repeated this calculation for different
values of $\delta$ for each value of $h$; for every value of
$\delta$ the curve $m_h(T)$ was averaged over 25 independent runs.
Typical curves $m_h(T)$ are shown in Fig.\ref{fig5} for different
values of $h$ and $\delta$. From these curves we estimated the
stability lines $T_s(\delta)$ (i.e., the temperature above which
$m_h=0$) for $h = 1,...,5$ and \mbox{$0<\delta<4.1$}. The
stability lines are shown in Fig.\ref{fig4}

As in the MF case, we observe the existence of regions below
$T_c^{(2)}$  where no stable striped solutions exist, at least up
to $\delta=3$; for values of $\delta>3$ the data becomes very
noisy (probably due to finite size effects) and the large error
bars do not allow to identify clearly those regions for the
present system sizes. We found that the equilibrium phase in those
regions is a nematic one, instead of an hybrid state, as in the MF
case (we checked the possible presence of hybrid states in those
regions and they always decay after a short time into a nematic
one). The nematic phase is characterized by an algebraic decay of
the spatial spin-spin correlations in one of the coordinate
directions and an exponential decay in the other. This can be
study through the structure factor (Fourier transform of the
correlation function):

\begin{equation}
  {\cal S}(\vec{k}) \equiv \left\langle \left| \hat{S}_{\vec{k}} \right|^2
  \right\rangle
\end{equation}

\noindent where $\hat{S}_{\vec{k}} = \frac{1}{N} \, \sum_i{S_i \,
e^{-i \vec{k}\vec{r}_i}}$. Cannas {\it et al}\cite{CaMiStTa2006}
calculated an approximate expression for the nematic phase
structure factor:

\begin{equation}
  \label{Sk}
  {\cal S}(\vec{k}) \approx \frac{\delta_{k_y,0}}{2\sqrt{N}} \,
  \left( \frac{\lambda}{(k_x-k_0)^2 + \lambda^2} +
         \frac{\lambda}{(k_x+k_0)^2 + \lambda^2}  \right)
\end{equation}

\begin{figure}
\begin{center}
\includegraphics[scale=0.3,angle=0]{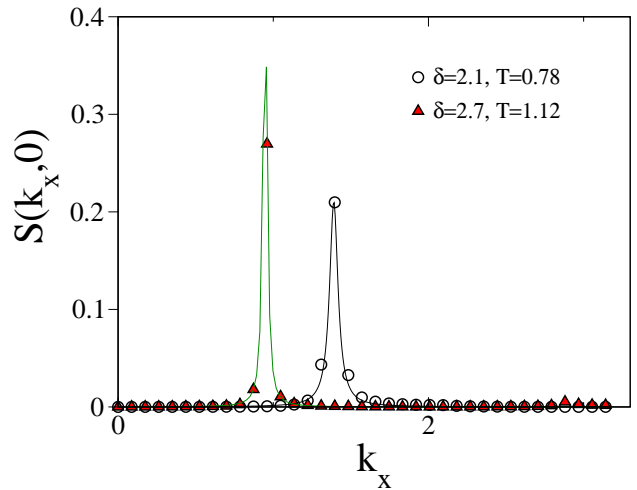}
\caption{\label{fig6} Structure factor for a $L=72$ system with \mbox{$\delta=2.1, T=0.78$,}
  ($\circ$) and $\delta=2.7, T=1.12$ ($\diamond$). Continuous lines correspond to
  fittings using (\ref{Sk}): for $\delta=2.1,T=0.78$, we obtained $\lambda=0.033, k_0=1.39$,
  for $\delta=2.7,T=1.12$, $\lambda=0.0149, k_0=0.947$. $S(0,k_y)=0$ $\forall k_y$ in both cases.}
\end{center}
\end{figure}

\begin{figure}
\begin{center}
\includegraphics[scale=0.3,angle=0]{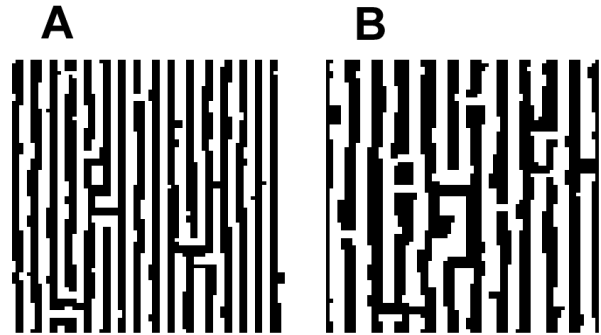}
\caption{\label{fig7} Typical nematic spin configurations for
$L=72$. ({\bf A})
                      $\delta = 2.10$, $T=0.78$; ({\bf B})
                      $\delta = 2.70$, $T=1.12$. }
\end{center}
\end{figure}

\noindent We ran simulations for $L=72$ and different $\delta$
values in the uncovered regions. The system was slowly heated from
zero temperature using the same process described above, up to a
temperature in the uncovered region, where we calculated ${\cal
S}(\vec{k})$ by averaging over $2 \times 10^5$ MCS. The typical
observed behavior of ${\cal S}(\vec{k})$ is shown in
Fig.\ref{fig6}, together with Lorentzian fittings using
Eq.(\ref{Sk}); two typical spin configurations in corresponding
regions can be seen in Fig.\ref{fig7} (compare with the results of
Ref.\cite{CaMiStTa2006}).

Following the same steps as in the MF case, we calculated next the
transition lines between different  phases at low temperature. The
transition lines were obtained by comparing the free energies of
the striped  phases of widths $h$ and $h+1$. The system size was
chosen in these calculations to be a multiple of both $2h$ and
$2(h+1)$. In what follows we will assume that the free energy of a
meta--stable state can be obtained by following a thermodynamical
path (that is, a close sequence of equilibrated states) from a
thermodynamically stable reference state. To calculate the free
energy of a striped phase of width $h$ we first computed the
internal energy per spin $u(T,\delta_i)\equiv \left< H\right>/N$
along a quasi-static path from an initially low temperature $T_0$
up to a working temperature $T<T_s(\delta_i)$ keeping $\delta_i$
constant and taking the initial spin configuration of a given
temperature as the final configuration of the previous one; the
value of $\delta_i$ was chosen well separated from the border
value at zero temperature between the striped phases of widths $h$
and $h+1$. The free energy was then obtained by
 numerically integrating the thermodynamic relation

 \begin{eqnarray}
\beta f_h(\beta,\delta_i)= \beta_0 \, f_h(\beta_0,\delta_i) +
\int_{\beta_0}^{\beta}u(\beta',\delta_i)d\beta',
\label{energia_livre_spin}
\end{eqnarray}

\noindent where $\beta =1/T$ and $\beta_0=1/T_0$. Once we arrive
to the final configuration at $(T,\delta_i)$ we perform a second
quasi-static path at constant temperature, by slowly changing
$\delta$, up to a final value $\delta_f$ corresponding to a
striped ground state of width $h+1$. Along this path we measure
the average exchange energy

\begin{equation}
u_{ex}(\beta,\delta) \equiv -\frac{1}{N}\left< \sum_{<i,j>} S_i
S_j \, \right>
\end{equation}

\noindent From the expression

\begin{equation}
  f = - \frac{1}{N \beta} \, \log{{\cal Z}}
  \label{ff}
\end{equation}

\noindent where ${\cal Z}$ is the partition function,  is easy to
see that

\begin{eqnarray}
  \frac{\partial f}{\partial \delta} &=&  u_{ex}(\beta,\delta).
\end{eqnarray}

\noindent Hence, the free energy along the last path can be
obtained by numerically integrating the equation

\begin{equation}
  f_h(\beta,\delta) =  \int_{\delta_i}^{\delta}  u_{ex}(\beta,\delta')\, d\delta' + f_h(\beta,\delta_i)
\end{equation}

\begin{figure}
\begin{center}
\includegraphics[scale=0.35,angle=-90]{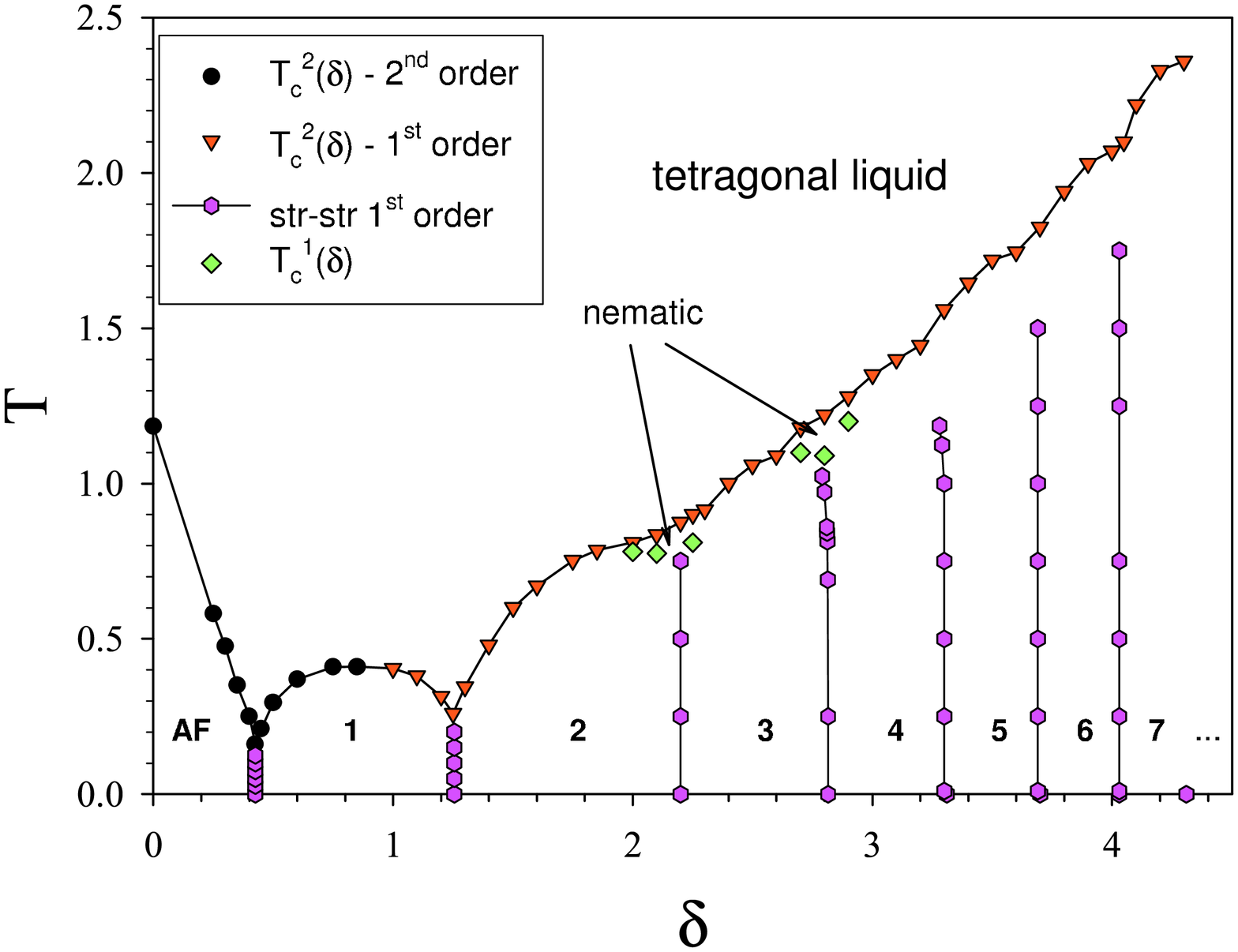}
\caption{\label{fig8} (Color on-line) Monte Carlo phase diagram.
The numbers indicate the equilibrium width of the low temperature
striped phases. The order of the different phase transition lines
are indicated in the inset (see text for details). The continuous
lines are a guide to the eye.}
\end{center}
\end{figure}

Repeating the same procedure for the striped phase $h+1$, but
following the second path in the inverse sense (i.e., decreasing
$\delta$ from $\delta_f$ down to $\delta_i$) we calculated the
free energy $f_{h+1}(\beta,\delta)$ at the same temperature. The
transition point $\delta_t(T)$ is obtained from the equation
$f_h(\beta,\delta_t)=f_{h+1}(\beta,\delta_t)$. We calculated the
first order transition lines between striped phases up to $h=6$.
The results are shown in the MC phase diagram of Fig.\ref{fig8}
(the transition lines between the AF and the $h=1$ and between the
$h=1$ and $h=2$ phases were already calculated in
Ref.\cite{GlTaCa2002}; we put it here for completeness). Notice
that all the calculated transition lines are almost independent of
$T$, at variance with the MF prediction.

\begin{figure}
\begin{center}
\includegraphics[scale=0.35,angle=0]{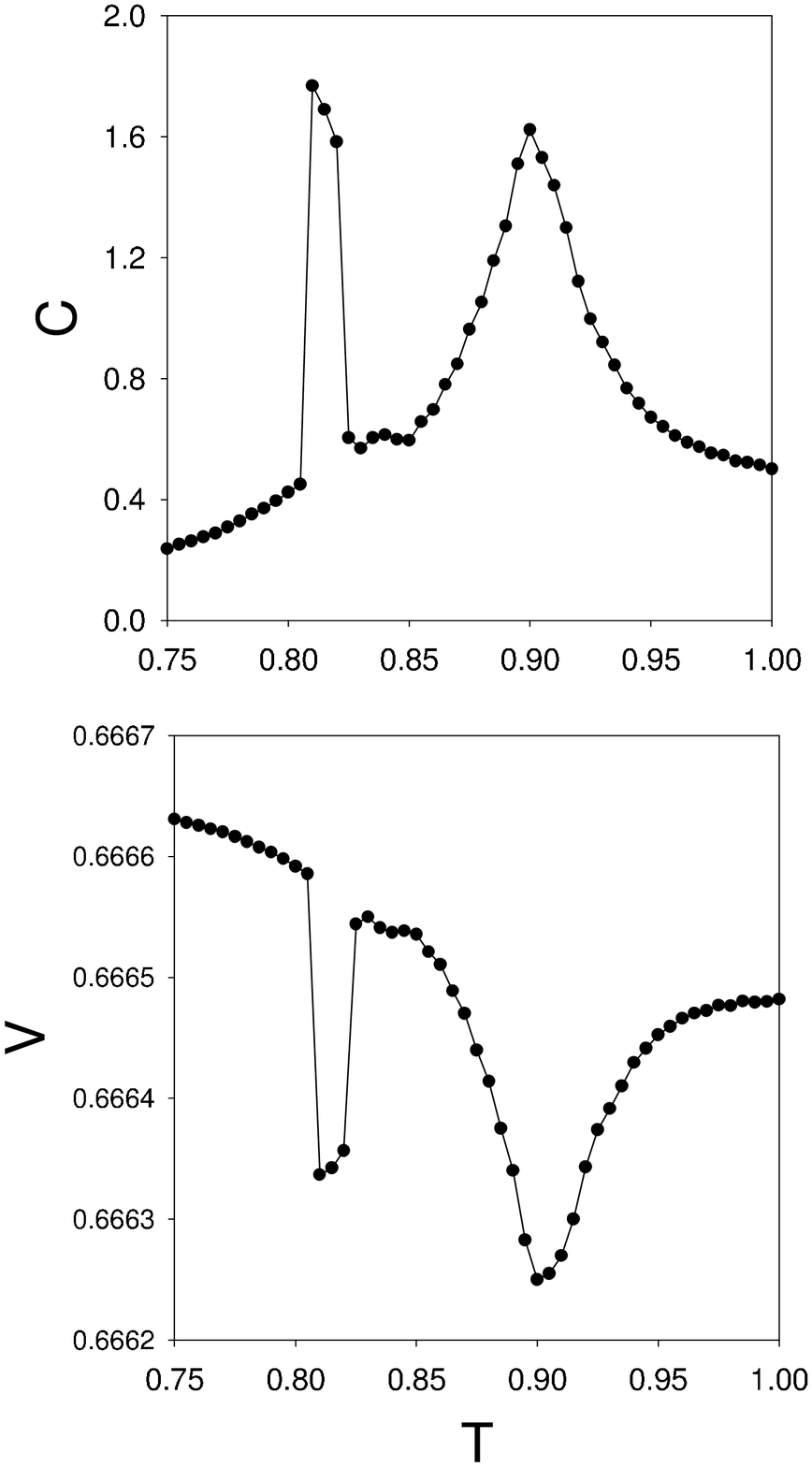}
\caption{\label{fig9} Specific heat and fourth order cumulant as a
function of $T$ for $\delta=2.25$ and $L=48$.}
\end{center}
\end{figure}

We next analyzed the transition temperature $T_c^{(1)}$ between
the nematic and the striped phases. In order to check whether the
stability lines $T_s(\delta)$ can be used to estimate $T_c^{(1)}$,
we analyzed the behavior of the specific heat and fourth order
cumulant Eqs.(\ref{heat}) and (\ref{cumulant}) around the regions
below $T_c^{(2)}$ where no striped states exist. The simulation
protocol used to determine $T_c^{(2)}$ is completely unable to
detect the low temperature transition at $T_c^{(1)}$. This is
because free energy barriers associated with both transitions for
the system sizes here considered are  larger around $T_c^{(1)}$
than around  $T_c^{(2)}$, as was shown in Ref.\cite{CaMiStTa2006}.
Indeed, a rough estimation of the average times needed by the
system to jump the free energy barrier between the striped and the
nematic phases are of the order of the millions of MC, thus
generating a strong meta-stability when the average times are of
the order of $10^5$ MCS\cite{CaMiStTa2006b}. In
Ref.\cite{CaMiStTa2006} it was shown that an accurate estimation
of $T_c^{(1)}$ for $\delta=2$ requires, for every temperature,
average times of the order of $2 \times 10^8$ MCS. However, we
verified that an average time of $5 \times 10^7$ is enough to
determine $T_c^{(1)}$ between the error bars we are using in the
present calculation (although such time scales are not enough to
determine the height of the specific heat maximum with precision
and therefore to allow a finite size scaling analysis). In order
to save computational effort, we used the following procedure for
fixed values of $\delta$ around the regions of interest: first we
ran the same simulation protocol as for $T_c^{(2)}$ down to low
temperatures and repeated it for the same parameter values, but
heating from a low temperature up to high temperatures and taking
as initial configuration the ordered striped state. In both cases
we calculated the internal energy $u(T)$ along the path. This
allowed us to determine the approximated location of  $T_c^{(1)}$,
by looking at the temperature range where the internal energy
exhibit meta-stability\cite{CaMiStTa2006b}. Then we calculated $C$
and $V$ for a limited set of temperatures in that region, by
taking averages for each temperature over a single MC run of $5
\times 10^7$ MCS. In order to get a more accurate estimation of
$T_c^{(2)}$ for the same values of $\delta$ we also repeated the
latter calculation for temperatures around the previous estimation
of $T_c^{(2)}$ taking averages over $10^7$ MCS. These calculations
were performed for $\delta=2.1$ and $2.25$ (near the $h=1 -h=2$
border with $L=48$. The behavior of $C$ and $V$ for $\delta=2.25$
 is shown in Fig.\ref{fig9} (compare with the results of
Ref.\cite{CaMiStTa2006}).  We verified that the location of the
low temperature peak of $C$ coincides between the error bars with
the value of $T_s(\delta)$ for the same values of $\delta$. The
values of  $T_c^{(1)}$ for $\delta=2$ (from
Ref.\cite{CaMiStTa2006}), $\delta=2.1,\, 2.25$ (from the above
calculation) and $\delta=2.7,\, 2.8,\, 2.9$ (estimated from the
stability lines) are shown by diamonds in the MC phase diagram of
Fig.\ref{fig8}. Reliable calculations of $T_c^{(1)}$ for larger
values of $\delta$ would require system sizes that are out of the
present computational capabilities.

\section{Discussion}
\label{disc}

 We have presented a detailed calculation of the
finite temperature phase diagram of the Ising dipolar model in the
range $0 \leq \delta \leq 4$, which allow  striped ground state
configurations of width up to $h=7$. We compared the predictions
of MF approximation with extensive MC numerical simulations.
Although the overall appearance of both phase diagrams looks
similar, several differences are remarkable.

The first difference to be noticed is the absence of nematic and
tetragonal order in the MF approximation. This results from the
fact that both phases are spatially disordered, which implies that
$\left\langle \hat{S}_{\vec{k}} \right\rangle =\hat{m}_{\vec{k}}
=0$ $\forall \vec{k}$. The characteristic features of those
states, namely, the broken rotational symmetry of the nematic
state and the discrete rotational symmetry of the tetragonal state
can only be observed when looking at the behavior of the spatial
correlations, or equivalently, of the structure
factor\cite{DeMaWh2000,CaMiStTa2006} Eq.(\ref{Sk}). Since
fluctuations are neglected in the MF approximation it follows that
${\cal S}(\vec{k})= \hat{m}_{\vec{k}} \hat{m}_{-\vec{k}}=0$ and
therefore the only possible spatially disordered solution within
this approximation is the paramagnetic one. On the other hand, the
MF approximation presents hybrid states solutions in the regions
of the phase diagram where MC predicts only nematic order.
Moreover, we verified that hybrid states are unstable in that
regions, suggesting that (in the language of renormalization
group) fluctuations play the role of a relevant scaling field that
turns the MF hybrid fixed points unstable towards nematic
attractors (in some sense, the hybrid states could be the closest
state to a nematic one that can obtained when fluctuations are
neglected). This would be consistent with the fact that
fluctuations, when included, can modify the continuous nature MF
prediction for the phase transition between the high temperature
disordered phase and the low temperature ordered one: Hartree
approximation applied to the continuous version of Hamiltonian
(\ref{Hamilton1}) predicts a fluctuation induced first order
transition {\it any finite} value of $\delta$~\cite{CaStTa2004},
which continuously fades out for increasing values of
$\delta$~\cite{CaMiStTa2006b}. In fact, MC simulations show a more
complex scenario, where the nature of the order--disorder phase
transition at $T_c^{(2)}$ depends on the value of $\delta$.

 Rastelli et al have shown that for $\delta=0$ the
 transition is indeed continuous and belongs to the universality
 class of the nearest neighbors Ising model\cite{RaReTa2006}. They also found a
 rather clear evidence of a second order transition for
 $\delta=0.85$, but with an unusual value for the critical exponent $\beta$\cite{RaReTa2006}. However, Cannas et al
 have shown that for $\delta=1$ the system presents a weak first
 order phase transition\cite{CaMiStTa2006}. These results are consistent with the
presence of a second order transition line for small values of
$\delta$, that joins with continuous slope a first order
transition line for larger values of $\delta$ at a tricritical
point somewhere between $\delta=0.85$ and $\delta=1$ and the
unusual critical exponents at $\delta=0.85$ is probably due to a
crossover effect near the tricritical point. There are also clear
evidences that the transition is first order for $\delta =2$
(Refs.\cite{CaStTa2004,CaMiStTa2006}) and $\delta=1.7,2.5$
(Ref.\cite{RaReTa2006}). The behavior of the fourth order cumulant
observed in the present work for $\delta=2.1$ and $\delta=2.25$ is
also consistent with a first order transition. For $\delta=3$ the
results of Rastelli et al\cite{RaReTa2006} appear to suggest that
the transition becomes continuous again. However, this is a matter
of debate\cite{CaMiStTa2006b} and numerical results using a
completely different technique\cite{CaDaRaRe2004} for
$\delta=4.45$ are also consistent with a first order transition.

We also presented numerical evidences of the presence of an
intermediate nematic phase between the disordered and the striped
ones in different parts of the phase diagram. Although it seems
that the nematic phase is only located near the border lines
between striped states, the presence of this phase in other
regions in narrow ranges of temperatures cannot be excluded and
larger system sizes should be required to clarify this.

 The existence of both type of scenarios for relatively large
values of $\delta$, namely one direct first order transition from
the striped phase to the tetragonal liquid or two phase
transitions with an intermediate nematic phase would be in
qualitative agreement with theoretical predictions based on a
continuous approach\cite{KaPo1993b,AbKaPoSa1995}. It is worth
mention that  Abanov et al\cite{AbKaPoSa1995} conjectured a second
order nematic-tetragonal phase transition; since their whole
analysis is based on mean field arguments, the disagreement with
the MC results can be understood from the fluctuation-induced
nature of the transition (see the discussion in
Refs.\cite{CaStTa2004,CaMiStTa2006}). Regarding the order of the
transition at $T_c^{(1)}(\delta)$, the situation is less clear.
Cannas et al\cite{CaMiStTa2006} have shown for $\delta=2$ that,
even the finite size scaling is consistent with a first order
transition, the energy changes continuously at $T_c^{(1)}(\delta)$
in the thermodynamic limit; this produces a saturation in the
associated specific heat peak, behavior that strongly resembles
that observed in a Kosterlitz-Thouless transition. That could be
indicative of the emergency of smectic order between the nematic
and the striped phases for larger system sizes and would be in
qualitative agreement with theoretical predictions based on a
continuous approach\cite{KaPo1993b,AbKaPoSa1995}. If that would be
the case, probably our calculation of $T_c^{(1)}(\delta)$
overestimates the true transition temperature, since it is known
that the specific heat peak  locates above the KT transition
temperature\cite{ChLu1995}, and therefore the nematic phase would
extend in larger regions of the phase diagram. However, at the
present level it is very difficult to improve this estimation due
to finite size effects.

Regarding the low temperature  behavior, a remarkable prediction
of  both MF and MC is the existence of an increasingly large
number of striped meta-stable states as $\delta$ increases.

Finally, we found that, at variance with the MF prediction, up to
$\delta=4$ the transition lines between striped phases are
completely vertical, imply temperature independence of the stripe
width. This suggests the existence of some large threshold value
of $\delta$, above which wall fluctuations makes the system to
cross over to a "mean field regime", where it starts to exhibit
temperature dependency of the stripe width.

 Fruitful discussions with D. A. Stariolo, F. A. Tamarit,
P. M. Gleiser, D. Pescia,  A. Vindigni and O. Portmann are
acknowledged. This work was partially supported by grants from
CONICET (Argentina), SeCyT, Universidad Nacional de C\'ordoba
(Argentina) and ICTP grant NET-61 (Italy).

\appendix
\section{}

\label{AppA}

When $k_y=0$ Eq.(\ref{fourierJ}) can be written as

\begin{equation}
  \hat{J}(k_x,0) =  2\, \delta (\cos{k_x}+1) - S(k_x,0)
  \label{Jkxapp}
\end{equation}

\noindent with

\begin{equation}
  S(k_x,0) \equiv \sum_i \frac{1}{r_{ij}^3} \cos{(k_x x_i)}
\end{equation}

\noindent where $x_i$ is the $x$ component of $\vec{r}_i$.
This last term can be rewritten as follows:

\begin{equation}
  S(k_x,0)  = 2 \sum_{x=1}^{\infty} \, \cos{(k_x x)}  \: R(x) + 2 \, \zeta(3)
  \label{Skx}
\end{equation}

\noindent with

\begin{eqnarray}
  R(x) \equiv \sum_{y=-\infty}^{\infty} \frac{1}{{(x^2+y^2)^{\frac{3}{2}}}}
\end{eqnarray}
\begin{eqnarray}
  \zeta(3) &\equiv& \sum_{y=1}^{\infty} \frac{1}{y^3} \approx 1.202
\end{eqnarray}

\noindent where $x,y$ represents the Cartesian coordinates of each
site and $\zeta(x)$ is the Riemann zeta-function. $R(x)$ can be
approximated by\cite{CzVi1989}

\begin{eqnarray}
  R(x) \approx \int_{-\infty}^{\infty}
  \frac{dy}{{(x^2+y^2)^{\frac{3}{2}}}} = \frac{2}{x^2}
\end{eqnarray}

\noindent which has an error of $\%1$ in the worst case $(x=1)$.
Inserting this in (\ref{Skx}) we get

\begin{eqnarray}
  S(k_x,0) &\approx& 4 \sum_{x=1}^{\infty}  \frac{\cos{(k_x x)}}{x^2}
  + \, 2 \, \zeta(3)  \nonumber \\
  &=& k_x^2 - 2 \pi |k_x| + \frac{2 \pi^2}{3} + 2 \, \zeta(3)
\end{eqnarray}

\noindent which replaced in Eq.(\ref{Jkxapp}) gives the expression
(\ref{Jkx}).

%\bibliography{ultrathin}

\end{document}